\documentclass[12pt,a4paper]{article}
\usepackage{amsfonts,latexsym}
\usepackage{amsmath,amssymb}
\usepackage{graphicx,color}

\renewcommand{\thefootnote}{\fnsymbol{footnote}}

\begin{document}

\vspace{12mm}

\begin{center}
{{{\Large {\bf Two instabilities of Schwarzschild-AdS black holes\\ in Einstein-Weyl-scalar theory  }}}}\\[10mm]

{Yun Soo Myung\footnote{e-mail address: ysmyung@inje.ac.kr}}\\[8mm]

{Institute of Basic Sciences and Department  of Computer Simulation, Inje University, \\ Gimhae 50834, Korea\\[0pt] }

\end{center}

\vspace{2mm}

\begin{abstract}
Stability of Schwarzschild-AdS (SAdS) black
hole is investigated  in  Einstein-Weyl-scalar (EWS) theory with a negative cosmological constant. Here, we introduce a quadratic scalar coupling to the Weyl term, instead of the Gauss-Bonnet term.
The linearized EWS theory  admits the Lichnerowicz equation for Einstein tensor as well as  scalar equation.
The linearized Einstein-tensor carries with a regular mass term (${\cal M}^2$), whereas the linearized scalar has a tachyonic  mass term ($-3r_0^2/m^2r^6$).
Two instabilities   of SAdS black hole in EWS theory  are found as  Gregory-Laflamme  and  tachyonic instabilities.
It shows that the  correlated stability conjecture holds  for small SAdS black holes obtained  from EWS theory by establishing  a close relation between Gregory-Laflamme and  thermodynamic instabilities.
On the other hand, tachyonic instability of SAdS black hole can be used for making five branches of  scalarized black holes  when considering  proper thermodynamic quantities of EWS theory (${\cal M}^2>0$).

\end{abstract}
\vspace{5mm}

\newpage
\renewcommand{\thefootnote}{\arabic{footnote}}
\setcounter{footnote}{0}


\section{Introduction}

It is well known that the dynamical stability of asymptotically flat black holes obtained from Einstein gravity  confirms the existence of these black holes in curved spacetimes~\cite{chan}.
If a solution of the black hole is dynamically unstable, it is no longer  considered as a truly black hole.
For Schwarzschild black hole, the Regge-Wheeler prescription works to indicate  dynamical stability even though it is  thermodynamic unstable in canonical ensemble (CE)~\cite{Regge:1957td,Zerilli:1970se}.
It was shown that the  Reissner-Nordstr\"{o}m (RN) black hole is stable against the tensor-vector perturbations~\cite{Zerilli:1974ai,Moncrief:1974ng}, while its thermodynamic (in)stabilities are found in CE.
The Kerr black hole is stable against the gravitational perturbations~\cite{Press:1973zz}, whereas its thermodynamic (in)stabilities are found in CE. This implies that there is no connection between dynamical instability and  local thermodynamic instability for asymptotically flat  black holes. As was pointed out in~\cite{Cardoso:2001bb,Ishibashi:2003ap,Moon:2011sz}, the SAdS black hole is dynamically stable against metric perturbations when adopting Regge-Wheeler  prescription, even though its thermodynamic (in)stabilities are found in CE.

However, there was  a close connection between
dynamical instability and  local thermodynamic instability for the black strings/branes in CE~\cite{Gubser:2000ec}.  This Gubser-
Mitra proposal was known to be the correlated stability conjecture (CSC)~\cite{Harmark:2007md} which  states  clearly that   gravitational systems with translational symmetry and infinite extent exhibit Gregory-Laflamme (GL) instability~\cite{Gregory:1993vy}, if and only if they have  local thermodynamic instability in CE.   Here, GL instability implies the dynamical instability. It is clear that the CSC does not hold for Schwarzschild and  SAdS black holes found in Einstein gravity  with a negative cosmological constant because they have no such  translational symmetry
and infinite extent.

 At this stage, we briefly explain the GL instability of the five-dimensional (5D) black string. Considering four-dimensional (4D) metric perturbation $h_{\mu\nu}^{(4)}=e^{\Omega t}e^{ik_z z}H_{\mu\nu}$
 around the 5D black string $ds^2_{(5)}=ds^2_{(4)}+dz^2$ background, its linearized Einstein equation leads to $(\bar{\Delta}+k_z^2)h_{\mu\nu}^{(4)}=0$ with $\bar{\Delta}$ 4D Lichnerowicz operator. The GL instability is an $s$-wave ($l=0$) spherically symmetric instability from 4D perspective. Solving $H_{tr}$-equation,  one found that the  GL instability bound is given by $0<k_z<0.876/r_+$ which is a long wavelength instability~\cite{Gregory:1993vy}.
In addition, the 4D dRGT  massive gravity~\cite{deRham:2010kj} which has a Schwarzschild solution when formulated in a diagonal bimetric form is subject to a direct analogous $s$-wave instability~\cite{Babichev:2013una,Brito:2013wya}. Furthermore,  a direct analogous $s$-wave instability bound for Schwarzschild black hole  was found as $0<m<0.876/r_+$ in Einstein-Weyl (Ricci quadratic) gravity when solving the linearized Einstein equation is given by $(\bar{\Delta}+m^2)\delta R_{\mu\nu}=0$~\cite{Myung:2013doa}. Two linearized Einstein equations become the same and thus, the same instability bound is obtained when replacing $k_z^2$ and $h_{\mu\nu}^{(4)}$ by $m^2$ and $\delta R_{\mu\nu}$~\cite{Stelle:2017bdu}.
Hence, it suggests that   the direct analogous $s$-wave instability is regarded  as the GL instability even though there is no direction with translation symmetry.

 Similarly, one expects that another version of CSC  holds for small SAdS black hole in higher curvature gravity with a negative cosmological constant because translational symmetry (higher dimensions) might be handled  by the massiveness (higher curvature gravity). Here, the negative cosmological constant is necessary to have its thermodynamic (in)stabilities of black hole in CE.
 Fortunately,  it was found that another version of CSC holds for small SAdS black holes found from Einstein-Weyl gravity with a negative cosmological constant  by establishing a close relation between GL instability and local thermodynamic instability in CE~\cite{Myung:2013uka}. Also, this CSC holds for  small SAdS black holes obtained from  Einstein-Ricci cubic gravity with a negative cosmological constant~\cite{Myung:2018ete,Myung:2018qsn}. For simplicity, we call another version of CSC as CSC in the present work.
 In these higher curvature gravity theories with negative cosmological constant, it is obvious that small (large) black holes are determined  precisely by negative (positive)  heat capacities  in CE and   the 5D black sting $(k^2_z)$ is replaced by the massiveness ($m^2\not=0$) of a massive spin-2 mode. In this case,  a massive spin-2 mode may be  described by either the linearized  Einstein tensor $\delta G_{\mu\nu}$ or metric tensor $h_{\mu\nu}$.

On the other hand, recently, there was a significant progress in obtaining black holes with scalar hair via spontaneous scalarization. This indicates  an example for evasion of no-hair theorem.
Here,  the tachyonic instability of linearized scalar propagating around bald black holes is regarded as  a hallmark for emerging  scalarized black holes when introducing a quadratic scalar coupling to the Gauss-Bonnet term~\cite{Doneva:2017bvd,Silva:2017uqg,Antoniou:2017acq}.
So far, the Einstein-Gauss-Bonnet-scalar (EGBS) theory is considered  as  a simple model to induce  tachyonic instability because the Gauss-Bonnet term is a topological term in four dimensions.
In its linearized theories, it is important to note that  the dynamical instability of bald black holes   is  determined only by   the linearized scalar equation because its linearized Einstein equation  is the same as that obtained from Einstein gravity which was shown  be stable tensor  modes~\cite{Myung:2018iyq}.  So, it is  clear that the dynamical instability of bald  black holes  is determined  by  tachyonic instability for scalar but its purpose is quite different from the conventional aspects. We note here that tachyonic instability  has nothing to do with local thermodynamic instability, but it is regarded as  a hallmark for emerging infinite branches of scalarized black holes.

In this work, we wish to investigate the stability of SAdS black
hole in  EWS theory with a negative cosmological constant and  a quadratic scalar coupling to the Weyl term. This model suggests a combined picture for establishing the CSC and generating scalarized black holes.
The linearized theory  admits the Lichnerowicz equation for Einstein tensor as well as  scalar equation.
The linearized Einstein-tensor carries with a regular mass term (${\cal M}^2=m^2-2/\ell^2$), whereas the linearized scalar has a tachyonic  mass term ($-3r_0^2/m^2r^6$).
Two instabilities   of SAdS black hole in EWS theory  are described  by  GL and  tachyonic instabilities.
We check that the  CSC holds  for small SAdS black holes obtained  from EWS theory by confirming  a close relation between GL instability and  local thermodynamic instability in CE.
On the other hand, tachyonic instability of SAdS black hole will be  used for making five branches of  scalarized black holes when preferring proper thermodynamic quantities.

\section{EWS theory and its black hole thermodynamics } \label{sec1}

The  EWS theory with a negative cosmological constant $\Lambda<0$ is given by
\begin{equation}
S_{\rm EWSc}=\frac{1}{16 \pi}\int d^4 x\sqrt{-g}\Big[ R-2\Lambda -2\partial_\mu \phi \partial^\mu \phi-\frac{f(\phi)}{2m^2} C^2\Big],\label{Action}
\end{equation}
where $f(\phi)=1-\phi^2$ is  a quadratic scalar coupling function, $m^2$ denotes a mass coupling parameter, and $C^2$ represents the Weyl term  given  by
\begin{equation}
C^2(\equiv C_{\mu\nu\rho\sigma}C^{\mu\nu\rho\sigma})=2(R_{\mu\nu}R^{\mu\nu}-\frac{R^2}{3})+{\cal R}_{\rm GB}^2.\label{Action2}
\end{equation}
Here,  ${\cal R}_{\rm GB}^2$ is  the Gauss-Bonnet term defined by $R^2-4R_{\mu\nu}R^{\mu\nu}+R_{\mu\nu\rho\sigma}R^{\mu\nu\rho\sigma}$.
A this stage, it is worth noting  that  scalar couplings to Gauss-Bonnet term  were mostly used to find scalarized black holes within EGBS theory because it provides a tachyonic mass term for a linearized scalar without modifying  the linearized Einstein equation~\cite{Doneva:2017bvd,Silva:2017uqg,Antoniou:2017acq}. This is so because the Gauss-Bonnet term is a topological term in four dimensions.

Varying the action (\ref{Action}) with respect to metric tensor, we derive  the Einstein  equation
\begin{eqnarray}
 G_{\mu\nu}= 2\partial _\mu \phi\partial _\nu \phi -(\partial \phi)^2g_{\mu\nu}+\frac{2(1-\phi^2)B_{\mu\nu}}{m^2}-\frac{\Gamma_{\mu\nu}}{m^2}, \label{equa1}
\end{eqnarray}
where $G_{\mu\nu}=R_{\mu\nu}-(R/2)g_{\mu\nu}+\Lambda g_{\mu\nu}$ is  the Einstein tensor.
Here, $B_{\mu\nu} (B^\mu~_\mu=0)$ coming  from the first part of $C^2$ in Eq.(\ref{Action2}) is the Bach tensor given by
\begin{eqnarray}
B_{\mu\nu}&=& R_{\mu\rho\nu\sigma}R^{\rho\sigma}-\frac{g_{\mu\nu}}{4} R_{\rho\sigma}R^{\rho\sigma}- \frac{R}{3}\Big(R_{\mu\nu}-\frac{g_{\mu\nu}}{4}R\Big) \nonumber \\
&+& \frac{1}{2}\Big(\nabla^2R_{\mu\nu}-\frac{g_{\mu\nu}}{6}\nabla^2 R-\frac{1}{3} \nabla_\mu\nabla_\nu R\Big) \label{bach}
\end{eqnarray}
and  $\Gamma_{\mu\nu}$   is given by
\begin{eqnarray}
\Gamma_{\mu\nu}&=&-\frac{2}{3}R\nabla_{(\mu} \Psi_{\nu)}-2\nabla^\alpha \Psi_\alpha \Big(R_{\mu\nu}-\frac{g_{\mu\nu}}{3}R\Big)+ 4R_{(\mu|\alpha|}\nabla^\alpha \Psi_{\nu)} \nonumber \\
&-&4 R^{\alpha\beta}\nabla_\alpha\Psi_\beta g_{\mu\nu}
+4R^{\beta}_{~\mu\alpha\nu}\nabla^\alpha\Psi_\beta  \label{equa2}
\end{eqnarray}
with
\begin{equation}
\Psi_{\mu}= 2\phi \partial_\mu \phi.
\end{equation}
Its trace is not zero as  $\Gamma^\mu~_\mu=R\nabla^\rho\Psi_\rho-2R^{\rho\sigma}\nabla_\rho\Psi_\sigma$.

Importantly, the scalar  equation takes the form
\begin{equation}
\nabla^2 \phi +\frac{C^2}{4m^2} \phi=0 \label{s-equa}.
\end{equation}

Considering the vanishing scalar $\bar{\phi}=0$,  the SAdS black hole  solution is found  from Eqs.(\ref{equa1}) and (\ref{s-equa}) as
\begin{equation} \label{ansatz}
ds^2_{\rm SAdS}= \bar{g}_{\mu\nu}dx^\mu dx^\nu=-\Big(1-\frac{r_0}{r}+\frac{r^2}{\ell^2}\Big)dt^2+\frac{dr^2}{\Big(1-\frac{r_0}{r}++\frac{r^2}{\ell^2}\Big)}+r^2d\Omega^2_2
\end{equation}
with AdS curvature radius $\ell^2=-3/\Lambda$ and a mass parameter  of SAdS black hole $r_0=2M$.
In this case, the horizon radius $r_+$ is the largest root  to $1-\frac{r_0}{r}+\frac{r^2}{\ell^2}=0$.
It is determined by
\begin{equation}
r_+=\frac{-2 \cdot 3^{1/3}\ell^2+(2\ell^4)^{1/3}\Big(9r_0+\sqrt{3(4\ell^2+27r_0^2)}\Big)^{2/3}}{(6\ell)^{2/3}\Big(9r_0+\sqrt{3(4\ell^2+27r_0^2)}\Big)^{1/3}}.
\end{equation}
A connection between $r_0$ and $r_+$ takes the form of $r_0=r_+(1+r_+^2/\ell^2)$.
This SAdS black hole background yields  $\bar{R}_{\mu\nu\rho\sigma}\not=0,~\bar{R}_{\mu\nu}=\Lambda\bar{g}_{\mu\nu},$ and $\bar{R}=4\Lambda$.
In this case, we find  that  $\bar{C}^2=\frac{12r_0^2}{r^6}$ and $\bar{\cal R}^2_{\rm GB}=\frac{12r_0^2}{r^6}+\frac{8\Lambda^2}{3}$.
The Hawking temperature of this black hole  is given by
\begin{equation}
T_{\rm H}=\frac{1}{4\pi r_+}\Big(1+\frac{3r_+^2}{\ell^2}\Big). \label{HT}
\end{equation}
On the other hand, using the Abbott-Deser-Tekin method~\cite{Abbott:1981ff,Deser:2002rt}, Einstein-Weyl thermodynamic quantities of mass, heat capacity, and  Wald entropy are given by~\cite{Myung:2013uka,Myung:2018ete}
\begin{eqnarray} \label{ADT-t1}
M_{\rm EW}(m^2,r_+,\ell)&=&\frac{{\cal M}^2}{m^2}M(r_+,\ell),~C_{\rm EW}(m^2,r_+,\ell)=\frac{{\cal M}^2}{m^2}C(r_+,\ell),\\
 S_{\rm EW}(m^2,r_+,\ell)&=&\frac{{\cal M}^2}{m^2}S_{\rm BH}(r_+),\label{ADT-t2}
\end{eqnarray}
where the thermodynamic quantities for SAdS black holes in Einstein gravity take the forms
\begin{eqnarray}\label{Ein-t}
M(r_+,\ell)=\frac{r_+}{2}\Big(1+\frac{r_+^2}{\ell^2}\Big),~\quad C(r_+,\ell)=2\pi r_+^2\Big(\frac{3r_+^2+\ell^2}{3r_+^2-\ell^2}\Big),\quad S_{\rm BH}(r_+)=\pi r_+^2.
\end{eqnarray}
Here, the mass squared ${\cal M}^2$ of a massive spin-2 mode is given by
\begin{equation}
{\cal M}^2=m^2-\frac{2}{\ell^2},
\end{equation}
which is positive/negative for $m\gtrless \sqrt{2}/\ell$.
\begin{figure*}[t!]
   \centering
  \centering
  \includegraphics{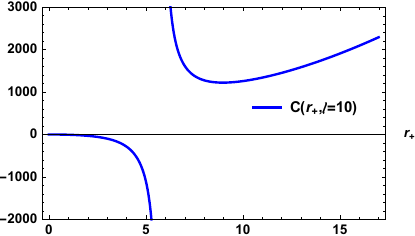}
  \hfill%
  \includegraphics{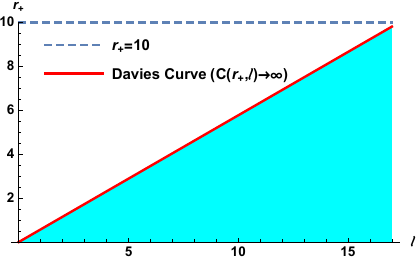}
\caption{ (Left) Heat capacity $C(r_+,\ell=10)$ for SAdS black hole. Heat capacity blows up at $r_+=r_*=5.773$ (red line), it is negative for $r_+<r_*$, and it is positive for $r_+>r_*$. This picture persists to $C_{\rm EW}$ for  ${\cal M}^2>0$. (Right) Davies curve for $C(r_+,\ell)\to\infty$. A red dot represents a red line $(\ell=10,r_+=r_*)$ in (Left). Cyan (white) regions denote $C(r_+,\ell)<0(C(r_+,\ell)>0)$.}
\end{figure*}
It is checked that the first-law of thermodynamics is satisfied within EWS theory as
\begin{equation}
dM_{\rm EW}=T_{\rm H} dS_{\rm EW},
\end{equation}
as well as the first-law is indeed satisfied in Einstein gravity
\begin{equation}
dM=T_{\rm H}dS_{\rm BH},
\end{equation}
where $`d$' represents the differentiation with respect to the horizon radius $r_+$ only.
As was shown for the SAdS black holes~\cite{Prestidge:1999uq}, an Euclidean nonconformal  negative mode which was originally proposed by Gross-Perry-Yaffe~\cite{Gross:1982cv} ceases to exist exactly when the heat capacity becomes positive.
This confirms the conjecture of Hawking and Page~\cite{Hawking:1982dh} and
demonstrates the correspondence between this eigenvalue spectrum and
the local thermodynamic stability of SAdS black hole in CE.

Hereafter, we consider the ${\cal M}^2>0$ case that is  described dominantly by the Einstein gravity. Observing Eq.(\ref{ADT-t1}), this case corresponds to having proper thermodynamic quantities.
The opposite  case of ${\cal M}^2<0$ corresponds to the case that is described  dominantly by the Weyl term and has improper thermodynamic quantities.
In the limit of $m^2\to 0/\infty$ with $\phi=0$, one recovers Weyl term (conformal gravity)/ Einstein gravity.
Since the heat capacity  $C(r_+,\ell=10)$ blows up at $r_+=r_*=\ell/\sqrt{3}=5.773\ell$ [see (Left) Fig. 1], one divides the black hole into small black hole with $r_+<r_*$ and  large  black hole with $r_+>r_*$.
This implies that the small black hole is thermodynamically unstable because $C_{\rm EW}<0~(C(r_+,\ell=10)<0)$, while the large black hole is thermodynamically stable because $C_{\rm EW}>0(C(r_+,\ell=10)>0)$.
Davies curve labels the positions where heat capacity diverges. According to the usual Ehrenfest classification, second-order phase transitions occur there.
(Right) Fig. 1 shows Davies curve representing for $C(r_+,\ell)\to\infty$.

\section{Two instabilities of SAdS black holes}
To perform the stability analysis of SAdS black hole in EWS theory, one needs   metric perturbation   $h_{\mu\nu}$ in ($g_{\mu\nu}=\bar{g}_{\mu\nu}+h_{\mu\nu}$) and scalar perturbation $\delta \phi$ in ($\phi=0+\delta\phi)$ propagating around (\ref{ansatz}).  Two linearized equations obtained  by linearizing Eqs.(\ref{equa1}) and (\ref{s-equa}) are decoupled  as~\cite{Lu:2011zk,Liu:2011kf,Myung:2013bow}
\begin{eqnarray}
 &&\Big(\bar{\Delta}_{\rm L}+\frac{6}{\ell^2}+{\cal M}^2\Big)\delta G_{\mu\nu}(h)=0, \label {lin-eq1}\\
 && \left(\bar{\Delta}_{\rm L}- \frac{3r_0^2}{m^2r^6}\right)\delta \phi= 0, \label{lin-eq2}
\end{eqnarray}
where the Lichnerowicz operator $\bar{\Delta}_{\rm L}$ is defined for tensor and scalar as
\begin{eqnarray}
&&\bar{\Delta}_{\rm L}\delta G_{\mu\nu}=-\bar{\nabla}^2\delta G_{\mu\nu}-2\bar{R}_{\mu\rho\nu\sigma}\delta G^{\rho\sigma}-\frac{6}{\ell^2} \delta G_{\mu\nu}, \label{ET-eq1} \\
&&\bar{\Delta}_{\rm L}\delta \phi=-\bar{\nabla}^2\delta \phi.  \label{ET-eq2}
\end{eqnarray}
The linearized Einstein tensor $\delta G_{\mu\nu}$ is defined by
\begin{equation}
\delta G_{\mu\nu}=\delta R_{\mu\nu}-\frac{\delta R}{2}\bar{g}_{\mu\nu}+\frac{3}{\ell^2}h_{\mu\nu} \label{lin-ET}
\end{equation}
with the linearized Ricci tensor $ \delta R_{\mu\nu}$ and the linearized Ricci scalar $\delta R$~\cite{Myung:2018ete}. Here, it is important to note  that `${\cal M}^2$' in  Eq.(\ref{lin-eq1}) is regarded as  a regular mass term for linearized Einstein tensor, while `$3r_0^2/m^2r^6$' in Eq.(\ref{lin-eq2}) is regarded as  a tachyonic mass term for linearized scalar. Also, we note that $\delta G^\mu~_\mu=-\delta R=0$ and the Bianchi identity $\bar{\nabla}^\mu \delta G_{\mu\nu}=0$ in the linearized EWS theory.

In addition, taking into account the transverse and traceless conditions of $\bar{\nabla}^\mu h_{\mu\nu}=0$ and $h^\mu~_\mu=0$, one finds~\cite{Myung:2013bow}
\begin{equation}
\delta G_{\mu\nu}=-\frac{1}{2}\Big(\bar{\Delta}_{\rm L}+\frac{6}{\ell^2}\Big)h_{\mu\nu}.\label{lin-eint}
\end{equation}
In this case, one rewrites Eq.(\ref{lin-eq1}) as a fourth-order equation
\begin{equation}
\Big(\bar{\Delta}_{\rm L}+\frac{6}{\ell^2}+{\cal M}^2\Big)\Big(\bar{\Delta}_{\rm L}+\frac{6}{\ell^2}\Big)h_{\mu\nu}=0 \label{lin-meq}
\end{equation}
which implies a linearized massless equation for $h_{\mu\nu}$
\begin{equation}
\Big(\bar{\Delta}_{\rm L}+\frac{6}{\ell^2}\Big)h_{\mu\nu}=0 \label{lin-m}
\end{equation}
and a linearized massive equation for $h^{{\cal M}}_{\mu\nu}$
\begin{equation}
\Big(\bar{\Delta}_{\rm L}+\frac{6}{\ell^2}+{\cal M}^2\Big)h^{{\cal M}}_{\mu\nu}=0.\label{lin-ma}
\end{equation}
We find that Eq.(\ref{lin-eq1}) is the same as Eq.(\ref{lin-ma}) when replacing $\delta G_{\mu\nu}$ by $h^{{\cal M}}_{\mu\nu}$.
Hence, we may use Eq.(\ref{lin-eq1}) as the linearized equation around SAdS black hole background for a massive spin-2 mode (linearized Einstein tensor).
However, one point to clarify  is that if one uses  Eq.(\ref{lin-ma}), one  may be  confronted  with the ghost issue because  Eq.(\ref{lin-ma}) arises from the fourth-order equation (\ref{lin-meq}).

\subsection{Gregory-Laflamme instability}
Eq.(\ref{lin-eq1}) leads to
\begin{equation}
\bar{\nabla}^2\delta G_{\mu\nu}+2\bar{R}_{\mu\rho\nu\sigma}\delta G^{\rho\sigma}-{\cal M}^2\delta G_{\mu\nu}=0.\label{slin-eq1}
\end{equation}
Actually, Eq.(\ref{slin-eq1}) describes a massive spin-2 mode ($\delta G_{\mu\nu}$) with mass ${\cal M}$  propagating on the  SAdS black hole background.
We wish to solve  equation (\ref{slin-eq1}) by adopting polar sector $\delta G_{\mu\nu}^{\rm e}(t,r)=e^{\Omega t}\delta G_{\mu\nu}(r)$. Its radial part  is initially  given by four  $\delta G_{tt}(r),~\delta G_{tr}(r),~\delta G_{rr}(r)$, and $\delta G_{\theta\theta}(r)$. Making use of $\delta G^\mu~_\mu=0$ and  $\bar{\nabla}^\mu \delta G_{\mu\nu}=0$, one derives one decoupled second-order equation for $s(l=0)$-mode $\delta G_{tr}$ as
\begin{equation}
A(r;r_0,\ell,\Omega^2,{\cal M}^2)\delta G_{tr}''(r)+B\delta G_{tr}'(r)+C\delta G_{tr}(r)=0, \label{A-eq}
\end{equation}
where the prime ($'$) denotes differentiation with respect to a  radial coordinate $r$. Three coefficients ($A,~B,~C$) were found in~\cite{Myung:2013uka}.
Solving Eq.(\ref{A-eq}) numerically  with appropriate boundary conditions, one reads off the GL instability bound from  Fig. 2 as
\begin{equation}
0<{\cal M}<{\cal M}^{\rm th}, \label{GL-bound}
\end{equation}
where ${\cal M}^{\rm th}=0.88/r_0=0.87$ for $r_0=1.01(r_+=1)$ denotes threshold mass of GL instability.

At this stage,
it is important to find that for $r_+=6>r_*=5.773$, the maximum value of $\Omega$ is less than $10^{-4}$, implying there is no unstable modes for large black hole with $r_+>r_*$.
The GL instability of small SAdS black holes sets in precisely when they are thermodynamically unstable ($C(r_+,\ell)<0$).
This means  that the CSC holds for  small SAdS black holes found in the EWS theory.
\begin{figure*}[t!]
   \centering
  \includegraphics{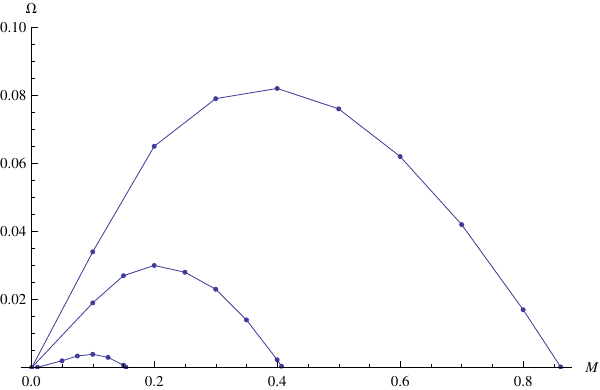}
\caption{ $\Omega$ graphs as function of mass parameter ${\cal M}$ for  linearized Einstein tensor $\delta G_{tr}$  with $r_+=1,~2,~4~(r_0=1.01,~2.08,~4.64)$ with $\ell=10$.
The thresholds of GL instability ($\Omega=0$) are located at ${\cal M}^{\rm th}=0.87,~0.42,~016$. }
\end{figure*}

Finally, it is worth noting that  we arrive at the same conclusion on the GL instability bound (\ref{GL-bound}) when using Eq.(\ref{lin-ma}), instead of Eq.(\ref{lin-eq1}).

\subsection{Tachyonic instability}
Tachyonic instability  has nothing to do with thermodynamic instability, but it is considered  as  an onset for emerging black holes with scalar hair.
From Eq.(\ref{lin-eq2}),
the  linearized scalar equation takes the form
\begin{equation}
 \Big(\bar{\nabla}^2 -\mu^2_{\rm S}\Big) \delta \phi= 0, \label{sads-eq}
 \end{equation}
 where a tachyonic mass squared is given by
 \begin{equation}
 \mu^2_{\rm S}=- \frac{3r_0^2}{m^2r^6}.
 \end{equation}
 Taking into account  separation of variables [$\delta \phi(t,r,\theta,\varphi)=\frac{u(r)}{r}e^{-i\omega t}Y_{lm}(\theta,\varphi)$]
and introducing a tortoise coordinate $r_*$ defined by $dr_*=\frac{dr}{1-2M/r+ r^2/\ell^2}$, the radial part of (\ref{sads-eq}) is given by
\begin{equation}
\frac{d^2u}{dr_*^2}+\Big[\omega^2-V_{\rm S}(r)\Big]u(r)=0,
\end{equation}
where the effective potential $V_{\rm S}(r)$  is
\begin{equation} \label{pot-c}
V_{\rm S}(r)=\Big(1-\frac{r_0}{r}+\frac{ r^2}{\ell^2}\Big)\Big[\frac{r_0}{r^3}+\frac{l(l+1)}{r^2}+\frac{2}{\ell^2}-\frac{3r_0^2}{m^2r^6}\Big].
\end{equation}
\begin{figure*}[t!]
   \centering
  \centering
  \includegraphics{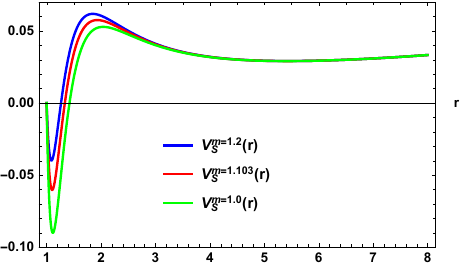}
  \hfill%
  \includegraphics{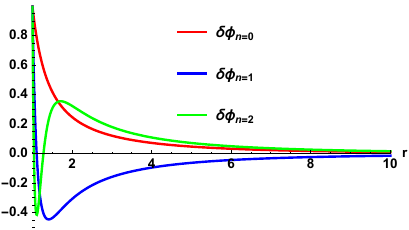}
\caption{(Left) Scalar potential $V_{\rm S}(r\in[r_+=1,10])$ for $s(l=0)$-mode $\delta \phi$  with $r_0=1.01$,  $\ell=10$, and $m=1.2$ (stable), $1.103(=m^{\rm th}),~1.0$ (unstable). The negative region increases as $m$ decreases.
(Right) Profile of $\delta \phi_n(r\in[r_+=1,10])$ with node number $n=0,~1,~2$ and $\ell=10$. }
\end{figure*}
For $V_{\rm S}(r)$ obtained from  EGBS theory, see Ref.~\cite{Guo:2020zqm,Zou:2023inv}.
 We could not impose  a sufficient condition of tachyonic instability because  $\int^{\infty}_{-\infty}V_{\rm S}(r)dr_*$ $\to \infty$ for any $\ell>0$.
 From (Left) Fig. 3, the negative region becomes wide and deep as the mass parameter $m$ decreases, implying tachyonic  instability of  SAdS black hole.
To determine the threshold of tachyonic instability for $s(l=0)$-mode $\delta \phi$, one  has to solve Eq.(\ref{sads-eq}) with $\omega=i\Omega$ directly,
which may  allow an exponentially growing mode of  $e^{\Omega t}$ as  an unstable mode.
The bound for  tachyonic instability for $r_+=1~(r_0=1.01)$ and $\ell=10$  is achieved for
 \begin{equation}
 0<m<m^{\rm th},\label{ta-b}
 \end{equation}
where the threshold mass  is given by $m^{\rm th}=1.103$.
If one requires ${\cal M}^2>0~(m>\sqrt{2}/\ell)$, the bound for tachyonic instability is modified as
\begin{equation}
 0.141<m<1.103.  \label{tt-b}
 \end{equation}

On the other hand,  we consider the static  scalar equation (\ref{sads-eq}) with $\omega=0$. Solving this equation, we obtain a finite spectrum of parameter $m$ : $m\in [1.103=m^{\rm th}$, 0.435, 0.271, 0.197, 0.155], which defines five branches of scalarized black holes: $n=0((0.141,1.103]),~n=1((0.141,0.435]),~n=2((0.141,0.271]),~n=3((0.141,0.197]),$ and $n=4(0.141,0.155)$. This is because the condition of ${\cal M}^2>0$ puts a lower bound on $m$ ($m>0.141$). Here, $n=0,~1,~2,~3,4$ are  identified with the number of nodes for $\delta \phi_n(r)$ profile [see (Right) Fig. 3].
Hence, we expect that five branches ($n=0,~1,~2,~3,~4$) of scalarized black holes  would be found when  solving Eqs.(\ref{equa1}) and (\ref{s-equa}) numerically.
However, this computation seems  to be difficult  because Eq.(\ref{equa1}) includes fourth-order derivatives and its Ricci scalar is not zero.

\section{Discussions}
In this work, we have investigated the instability of SAdS black
hole in  EWS theory with a negative cosmological constant and  a quadratic scalar coupling to the Weyl term.
This theory has provided a unified picture for establishing  the correlated stability conjecture (CSC) and generating infinite branches of scalarized black holes.
The linearized theory  admits the Lichnerowicz equation for Einstein tensor (metric tensor) as well as  scalar equation.
The linearized Einstein-tensor (metric tensor) carries with a regular mass term (${\cal M}^2$), whereas the linearized scalar has a tachyonic  mass term ($-3r_0/m^2r^6$).
Two instabilities   of SAdS black hole in EWS theory  are described  by  Gregory-Laflamme (GL) and  tachyonic instabilities.
The GL instability bound takes the form   $0<{\cal M}<0.87$ with ${\cal M}=\sqrt{m^2-2/\ell^2}$ for $r_+=1$ and $\ell=10$, while the tachyonic instability bound is given by $0.141<m<1.103$.
Furthermore, we have shown that the CSC holds  for small SAdS black holes with $r_+<r_*$ obtained  from EWS theory by confirming  a close connection between GL instability and  local thermodynamic instability in canonical ensemble.
On the other hand, it seems that tachyonic instability of SAdS black hole has nothing to do with thermodynamic instability. In the case of EGBS theory, tachyonic instability implies  infinite branches of scalarized black holes.  However,   number of branches of scalarized black holes is five  when taking into account  proper thermodynamic quantities of EWS theory (${\cal M}^2>0$).

\end{document}